\documentclass[submission,copyright,creativecommons,noncommercial]{eptcs}
\providecommand{\event}{Post-Proceedings --- ThEdu'17} % Name of the event you are submitting to
\usepackage{underscore}           % Only needed if you use pdflatex.

\usepackage[utf8x]{inputenc}
\usepackage{array}
\usepackage{url}

\makeatletter
\newcommand{\lam}	 {\mbox{\(\lambda\)}}

\usepackage{ucs}
\usepackage[T1]{fontenc}
\usepackage{graphicx}
\usepackage{amsmath}
\usepackage{amsthm}
\usepackage{subcaption}
\newtheorem{definition}{Definition}

\usepackage{listings}

\title{A Theorem Prover for Scientific and Educational Purposes}
\author{Mario Frank\thanks{\url{http://orcid.org/0000-0001-8888-7475}} \quad Christoph Kreitz
    \institute{University of Potsdam, Institute for Computer Science, Potsdam, Germany}
    \email{\{mafrank,kreitz\}@uni-potsdam.de}
}

\def\titlerunning{A Theorem Prover for Scientific and Educational Purposes}
\def\authorrunning{M. Frank \& C. Kreitz}

\begin{document}
\maketitle

\begin{abstract}
  We present a prototype of an integrated reasoning environment for educational
  purposes. The presented tool is a fragment of a proof assistant and automated
  theorem prover. We describe the existing and planned functionality of the
  theorem prover and especially the functionality of the educational
  fragment. This currently supports working with terms of the untyped lambda
  calculus and addresses both undergraduate students and researchers. We show
  how the tool can be used to support the students' understanding of functional
  programming and discuss general problems related to the process of building
  theorem proving software that aims at supporting both  research and
  education.
\end{abstract}

\section{Introduction}

Interactive and automated deduction are important concepts in university
education when teaching students how to reason about the correctness of software and
hardware. But usually, the teaching has to be limited to the theoretical aspects
of theorem proving as there are no proof tools suitable for
undergraduates. Well-known proof assistants like Coq \cite{ref:Coq:manual},
NuPRL \cite{ref:NuPRL:book-full}, Isabelle
\cite{ref:Isabelle:Nipkow:2002:IPA:1791547}, Agda
\cite{ref:Agda:Norell:2009:DTP:1481861.1481862} and Idris
\cite{ref:Idris:JFP:9060502} are implemented specifically for scientific
purposes and in our experience have a learning curve too steep for beginning students.

Before students can be exposed to advanced deduction topics relevant for
reasoning about hardware and software, such as higher-order logics or type
theory, they first need to understand the basics of computability, functional
programming, and constructive logic. Among others, they need to understand the
computational semantics of the (untyped) lambda calculus
\cite{ref:lambda:Bar:84}. While in principle existing proof assistants could be
used for this purpose, solving the issue of the steep learning curve would
require a lot of effort.  Thus there is need for proof assistants specifically
suited for the needs of undergraduate students.

In this article we present the educational fragment of a proof assistant that
supports the (untyped) lambda calculus and specifically addresses
students. After a brief overview of the existing tools and approaches we first
discuss the broader context of our tool by presenting the existing and planned
functionality of the larger theorem proving environment.  After that, we review
the essential basics of the untyped lambda calculus and show the various
features of our tool and how they can be used. Finally, we discuss the current
state and limitation of our tool as well as possible future extensions.

\section{Related Work}

There are already some efforts to create educational tools for teaching the
basics of deduction.  Peter Sestoft~\cite{DBLP:journals/entcs/Sestoft01} created
a tool that can
demonstrate different reduction strategies for the untyped lambda calculus.  It
supports multiple strategies and can display all intermediate steps executed by
the strategies.  Named terms (abbreviations) are also supported, which
introduces the possibility of modularisation. But users do not apply individual
reduction steps themselves and they do not have to apply $\alpha$-conversions
explicitly as the system does it automatically for them.  So the learning
outcome is limited. Furthermore, the tool is not stand-alone since a web server
is needed to use it.

The most elaborated lambda simulator is the Penn Lambda
Calculator, described in \cite{Penn-Lambda-Calculator}. It supports the typed lambda calculus and is specifically designed as educational environment for learning natural language semantics. It also supports the creation of exercises and a conference mode for
e-Learning.  Furthermore, it includes a feature to grade  solutions of
students.  The implementation in Java makes it platform-independent. The main
difference between the Penn Lambda Calculator and our approach is that our
approach is built on top of a theorem prover and aims at getting students used
to formal reasoning tools.

The lambda calculus tracer TILC~\cite{ref:TILC:Ruiz:2009:TIL:1594413.1594563} is
able to display lambda terms as trees. It can highlight bound and free
variables, the most inner redex and also mark subterms of a selected lambda
term. Where possible, it can reduce lambda terms to normal form and show all
steps of the reduction. But it does not terminate if a non-normalisable term
like $(\lambda x. x x) (\lambda x. x x)$ is given as input. Furthermore, the
tool seems to be out of maintenance since after 2009 and there is only an
executable for Windows.
There are also some online
tools\footnote{\url{https://people.eecs.berkeley.edu/~gongliang13/lambda/}}
\footnote{\url{https://www.easycalculation.com/analytical/lambda-calculus.php}}
for lambda term evaluation. While the first does not support named lambda
terms, the latter does. Both are capable of evaluating lambda terms
automatically but students do not have the chance to perform the reductions or
$\alpha$-conversions themselves. There also is no offline version of these tools.
The lambda calculators implemented by Joerg
Endrullis\footnote{\url{http://joerg.endrullis.de/lambdaCalculator.html}} and a
student of the University of
Sidney\footnote{\url{https://github.com/scyptnex/lambda-calculator}} are Java
based tools.  The former shows the lambda terms as graphs and allows $\beta$-reduction
via clicking and term manipulation via drag and drop. But the source code does
not contain a main routine, so the tool cannot be started successfully. The
latter one is text based and does the $\beta$-reduction automatically.
Carl Burch implemented multiple tools for educational purposes. One of the
tools\footnote{\url{http://www.cburch.com/proj/lambda/}} is a lambda simulator
both as online JavaScript version and offline Java version.  It supports named
terms and reduces the given term to normal form automatically, if possible.
Another non-graphical
implementation\footnote{\url{https://github.com/sgillespie/lambda-calculus}} of
a lambda simulator was written in Haskell.  The typed terms are evaluated
automatically, too.

\section{The Complete Framework for Theorem Proving}

As already pointed out in the introduction, our educational tool for the lambda
calculus is part of a larger project that aims at supporting both interactive
and automated theorem proving.
Current interactive theorem provers like Isabelle, Coq, NuPRL, Agda and
Idris are successfully used in both scientific and industrial contexts.
Although some interactive theorem provers contain subsystems to delegate proof obligations to automated theorem provers like Vampire\cite{voronkov:vampire}, E\cite{schulz:eprover}, iProver\cite{korovin:iprover} and
leanCoP\cite{otten:leancop}, those subsystems have various problems.
Sledgehammer\cite{ref:Sledgehammer} for example is used for many years as interface between Isabelle and theorem provers like Vampire. But the interfaces to the automated theorem provers are occasionally unstable and the results from the automated theorem provers are in some cases not usable.
The main problem here is a missing exchange format that is also enforced. While
a de facto standard exists with the TPTP (Thousands of Problems for Theorem Provers)\cite{sutcliffe:tptp} language that also
contains the solution language TSTP (Thousands of Solutions from Theorem Provers), this standards do not require the automated
theorem provers to return complete proofs. In many cases, the automated theorem
provers, especially the ones based on resolution or equational reasoning, do not
store all proof steps during proof search, which makes it hard or even
impossible to communicate all details.  For example, the use of the axioms of
equalities, i.e.~symmetry, reflexivity and transitivity, is subsumed by the
inference that the theory of equalities was used. But this information may not
be helpful for interactive theorem provers as the proof cannot be reconstructed
easily\cite{ref:Sledgehammer}.  In some cases this even leads to loss of termination during
reconstruction like in the case of the subsystem Metis\cite{hurd:metis} of
Isabelle.
Another problem is the variable replacement by skolemisation that is not
communicated. And even if an automated theorem prover communicates a complete
proof, that proof may not be useful for interactive theorem provers if the proof
is not valid in the underlying logic of these provers.

The theorem prover presented here explicitly aims at enforcing complete proofs.
Also, it is intended to provide a framework for prototyping automated proof
calculi while offering standard decision procedures like unification
(e.g. \cite{robinson:unification,mm:unification} and standard methods
for optimisation like term indexing\cite{colomb:clauseindexing}).
In a broader sense, the framework can be seen as a logical framework, but in
detail it is more than just that. While logical frameworks usually provide the
most atomic layer of logic, i.e.~the standard logical connectives (conjunction,
disjunction, negation, and implication) together with a calculus like the
sequent calculus\cite{gentzen:sequent}, our framework enriches the standard with
multiple layers for specialisation to the proof domain. In our framework,
connectives like equivalence, NAND and NOR are not only defined in terms of the
basic ones but can also be used as atomic connectives without a need for
unfolding their definition. This saves time during proof search and still
preserves the possibility of unfolding for the complete proof.

\begin{figure}[htb]
 \centering
 \includegraphics[scale=0.55]{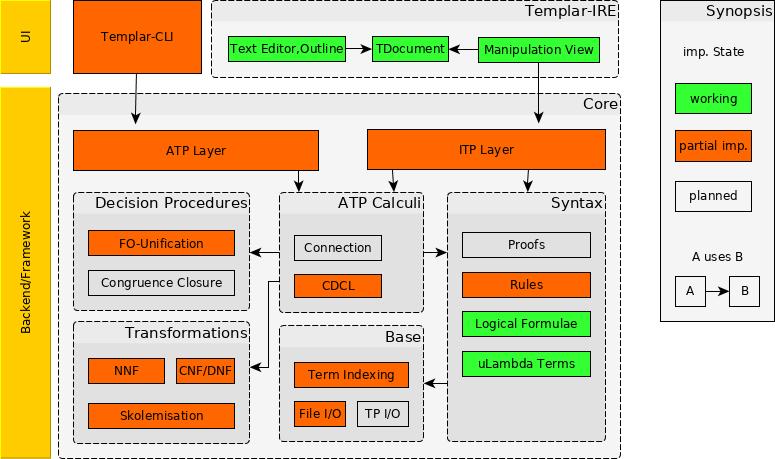}
 \caption{Architecture of the complete theorem proving environment}
 \label{figure:architecture}
\end{figure}

Those specialisations will enable our framework to select optimised calculi and
decision procedures for the problem or domain of interest.  To allow
specialisation while preserving completeness of information, a special
architecture is needed. A draft architecture is presented in figure
\ref{figure:architecture}.  The architecture of the system consists of the
framework and the UI layer.  The framework contains a set of decision
procedures, logical transformation procedures, syntax definitions for
computational terms, logical formulae along with inference rule and proof
schemes. Also, it has a base component with common utility methods for term
indexing and file I/O and also already existing prototype implementations of ATP
calculi. There are two interfaces that can be accessed externally, i.e.~the
interactive theorem proving (ITP) and the automated theorem proving (ATP)
layer. Both layers have direct access to the ATP calculi component to be able
providing its functionality. Moreover, the ITP layer also has direct access to
the syntax layer in order to provide it's functionality while the ATP layer
gains access to the syntax component via the ATP calculi component. The main
reason for the distinction in the syntax component access is that the ITP layer
must be able to handle all syntactical definitions directly as it communicates
directly with the (graphic) user interface and is not accessed from external
systems. The ATP layer on the other hand is planned to being accessible via
foreign function interfaces. The communication between the ATP component and
external systems is planned to be handled by the base component, especially the
theorem prover I/O module.

The most complex part is the syntax component as it needs to enable
specialisations. This component contains the untyped lambda syntax and rule
definitions that are currently used especially for the educational purpose.
Extending these definitions to type theoretical structures, such as the typed
lambda calculus would make a communication with ITP systems possible. Moreover,
there are implementations for logical connectives that shall be used for
automated theorem proving. Between those two definitions, a Curry-Howard-Style
transformation is planned,  which would make the results from the ATP system
reusable for interactive theorem proving.

We plan to provide access to our framework by strictly defined interfaces, like
foreign function interfaces. The main advantage of using foreign function
interfaces is that the communicated data do not need to be cached in files or
in streams but directly forward the syntactical structures to the target system
if the target system itself has some reasonably good interface. But as the
interfaces are not necessarily present, we focus on systems, where a foreign
function interface is easily establishable.

The candidates for interfacing are selected by specific criteria. First of
all, there must be a reasonable way to communicate with the theorem provers
without passing queries and results as strings via the shell. So, either foreign
function interfaces between the theorem, i.e.~the programming languages must
exist, or the theorem provers must have some structured network based
communication possibility, like webservices. This already makes NuPRL and Coq
good candidates as the former has network based interfaces and the latter one
has foreign function interfaces between OCaml, it's implementation language, and
C/C++. Moreover, those candidates are favoured above Isabelle as Isabelle
already has ATP connectivity, although it is text based.  For ATP systems, the
criterion would make E and Vampire good candidates as they are implemented in C
and C++ respectively which would make a communication even more easy. OCaml
based candidates would also be iProver and Leo\cite{benzmueller:leo3}.
As already the architecture shows, many parts of the system are currently in
planning phase or only exist as prototypes. Thus, we focus on the currently
usable parts, i.e.~the lambda calculus fragment and the IDE.

\section{Supported Syntax}

The syntax supported by our tool is an extension of the syntax of the untyped
lambda calculus.

\begin{definition}[Lambda Terms]
 The inductive definition of lambda terms is as follows:
 \begin{itemize}
  \item A variable \textbf{v} is a term.
  \item If \textbf{s} and \textbf{t} are terms, the application \textbf{s t} is
    a term.
  \item If \textbf{x} is a variable and \textbf{t} a term, then the abstraction \textbf{$\lambda x.t$} is a term.
  \item If \textbf{t} is a term, then $( t )$ is a term.
 \end{itemize}
\end{definition}

Applications are left associative and abstractions bind as far as possible.
$\alpha$-conversion substitutes a bound variable with a variable with a new name, i.e.
$\lambda x.t \, \overset{\alpha}{\rightarrow} \, \lambda x'.t[x/x']$ and
$\beta$-reduction reduces an application of the form $(\lambda x. t) s$ to
$t[x/s]$. In order to leverage the understanding of modularisation, we support
named terms, i.e.~abbreviations for complex lambda terms.  Thus, we add a rule
\ $t \, \overset{\equiv}{\rightarrow} \, def_{t}$ \ where $t$ must be a named
term and $def_{t}$ is the definition of the named term.

We also extend the syntax by support for multi-bindings, which, for instance,
allows to use  $\lambda x,y,z$ as
abbreviation for $\lambda x. \lambda y. \lambda z$.  Both the multi-binding
delimiter "$,$" and the normal binding delimiter "$.$" are configurable. For
instance, it is possible to define a whitespace as multi-binding delimiter. This
way the syntax can be made more conformant to the syntax of other proof
assistants like Coq where the whitespace delimits binding parameters. Moreover,
the grammar can support multiple variants to write a {\lam}, e.g. "\textbackslash"
and "\textbackslash{}lambda", which is also configurable.  The tool
automatically transforms both variants into the {\lam} symbol.  There is also
automatical renaming of the rules "\textbackslash{}alpha",
"\textbackslash{}beta" and "\textbackslash{}equiv" to $\alpha$, $\beta$ and $\equiv$
respectively.

It is possible to define both unnamed and named terms where the former have the
syntax \linebreak
 $\{ Term \, ( \rightarrow(\alpha | \beta | \equiv )\, Term )*\}$ and the
latter extend the unnamed term syntax with a leading $Name :=$ where the name
must begin with an upper case symbol. An example for a named term definition
would be \ $True \, := \, \{ \, \lambda x. \lambda y. x \,\}$.  So, named and
unnamed lambda terms are represented as list of terms, delimited by the rule
that was applied in order to generate them.

\section{Different Aspects of the Tool}

\begin{figure}[htb]
 \begin{center}
  \includegraphics[scale=0.48]{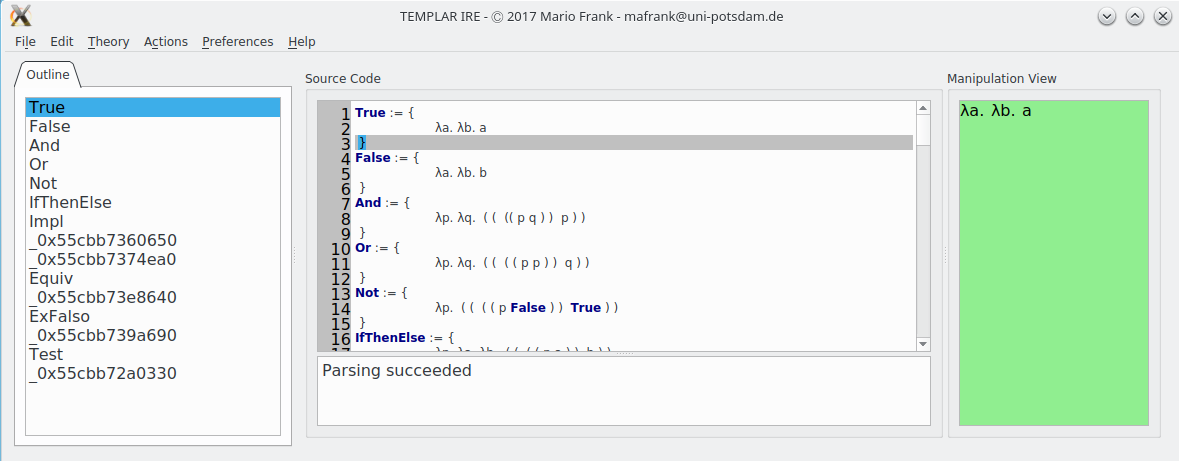}
  \caption{The user interface}
  \label{figure:UI}
 \end{center}
\end{figure}

The graphical user interface consists of three parts as shown in figure
\ref{figure:UI}.  The left sidebar, i.e.~the outline, shows the names or ids of
successfully parsed terms. The center contains the source code editor where the
lambda terms can be written. The right window, i.e.~the manipulation view, shows
the currently selected lambda term and its derived terms in a more interactive
variant.

Selecting the name or id of a term in the outline sets the lambda term as
content of the manipulation view and signals the code editor to move the cursor
to the end of the source code of the term.
We will focus on the functionality and use of the source code editor and the manipulation view.

\subsection{The Source Code Editor}

The source code editor is the main part of the integrated development resoning
environment and supports usual the functionality  common to most integrated
development environments.  For example, it supports syntax highlighting,
i.e.~showing the closing parenthesis for an opening parenthesis on the current
cursor position and highlighting the used named terms in blue if they are
defined and in red if they are not.

\begin{figure}[htb]
 \centering
 \includegraphics[scale=0.5]{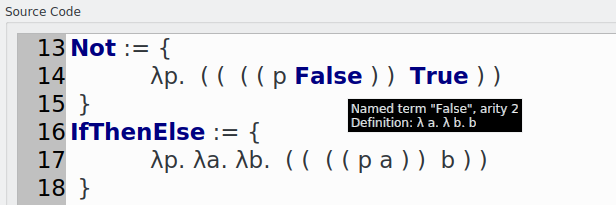}
 \caption{Tooltip with detailed information about the named term.}
 \label{figure:tooltip}
\end{figure}

Also, it is able to show the definition and arity of the named term as tool tip
when hovering over the name (see figure \ref{figure:tooltip}).  Moreover, the
source code editor supports dynamic code completion for named terms and
applicable rules ($\alpha$-conversion, $\beta$-reduction and named term expansion
$\equiv$). If a new named term is defined, it is automatically supported by the
code completion.
Writing in the text editor is more or less like in any other text editor except for
the automatic rewriting of special keywords. For example, when the user writes
"-$>$\textbackslash{}beta \quad \textbackslash{}lambda x. \textbackslash{}lambda y. f x y",
this is automatically rewritten to  $\text{-$>$}\beta \quad \lambda x. \lambda y. f x y$ while
the text is written. But this rewriting is only used for the presentation in the editor.
The file itself contains the written text in plain format, i.e. \textbackslash{}lambda.
Although the set of rewriting rules is currently fixed, 
it is in principle possible to extend it. Enabling users to extend the rules is
a part of the future work.
The source code editor does not yet support checking the
correctness of $\beta$-reduction, $\alpha$-conversion and $\equiv$-expansion that is
written. This is currently only supported by the manipulation view.

Below the source code editor, the parser output is shown. On error, the line and column of the erroneous part of the source code is shown.

\subsection{The Manipulation View}
The manipulation view shows the selected lambda term and its derived terms and
supports both informational actions and rule applications on the terms.

\subsubsection{Informational Actions}
Informational actions are unlocked by holding the control button pressed.  The
most relevant informational actions are the syntax highlighting. Here, clicking
on a parenthesis highlights the corresponding one, and clicking on the name of a
variable bound by an abstraction ($\lambda x$) highlights all occurrences of
this variable in the subterm.  Also, clicking on a bound variable somewhere in
the term highlights all occurrences of it and the location where it is bound,
i.e.~the abstraction.

Hovering over a named term reference shows its original definition of the named
term together with its arity as it is done in the source code editor. Here,
holding the control button pressed is not necessary.

\subsubsection{Rule Application}
The rule applications are the most important part of the manipulation view for
the students.  Rules can be applied only on the last term shown in the
manipulation view.
% alpha
\paragraph{$\alpha$-conversion}
The $\alpha$-conversion is triggered by double-clicking either on the variable name of
an abstraction or on the {\lam}-symbol itself.

\begin{figure}[htb]
 \centering
 \begin{subfigure}{0.49\textwidth}
    \includegraphics[scale=0.5]{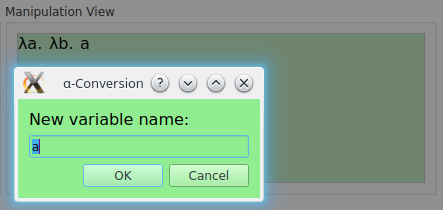}
    \caption{$\alpha$-conversion input}
    \label{figure:alpha}
 \end{subfigure}
 \begin{subfigure}{0.49\textwidth}
    \includegraphics[scale=0.5]{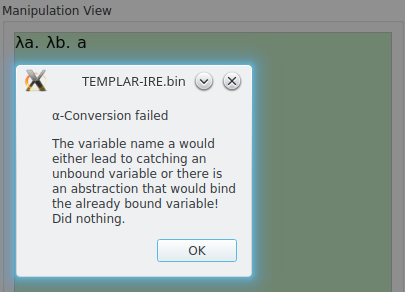}
    \caption{$\alpha$-conversion failure}
    \label{figure:alpha-fail}
 \end{subfigure}
 \caption{$\alpha$-conversion}
\end{figure}

This action opens opens an input field (see figure \ref{figure:alpha} where the
new name can be given for the variable. If the new name does not bind a
previously free variable or catches another bound variable, $\alpha$-conversion is
applied. Both the manipulation view and the source code are updated. Otherwise,
a warning message is shown (see figure \ref{figure:alpha-fail}).  The input
field also has to make sure that the variable naming conventions are preserved,
i.e.~no upper case variable names are allowed. A better way of applying the
$\alpha$-conversion whould be to make the double clicking on the variable switching to
edit mode and confirming the change with the return button.  This is subject for
improvements.

% beta
\begin{figure}[htb]
 \centering
 \begin{subfigure}{0.49\textwidth}
    \includegraphics[scale=0.6]{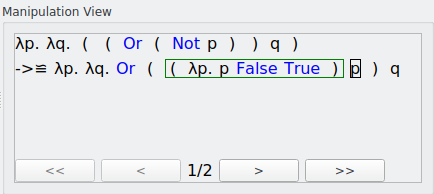}
    \caption{$\beta$-reduction choice 1}
    \label{figure:beta1}
 \end{subfigure}
 \begin{subfigure}{0.49\textwidth}
    \includegraphics[scale=0.6]{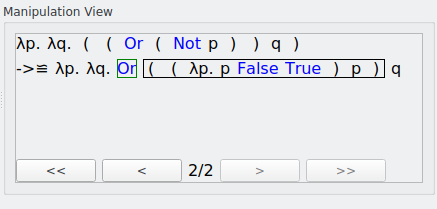}
    \caption{$\beta$-reduction choice 2}
    \label{figure:beta2}
 \end{subfigure}
 \caption{$\beta$-reduction choices}
\end{figure}

\paragraph{$\beta$-reduction}
If the last term is not in $\beta$-normal form, the $\beta$-reduction can be applied to
it. The manipulation view shows how many possible $\beta$-reductions are possible in a
navigation bar at the bottom. A $\beta$-reduction option is always shown by
highlighting the argument in a black box and the function where the argument can
be inserted in a green box (see \ref{figure:beta1}).  If there are multiple
possible reductions, the focus on the application to be reduced can be changed
with the shortcuts CTRL+P for previous and CTRL+N for next or with the
navigation buttons on the bottoms of the manipulation view (see figure
\ref{figure:beta1} and \ref{figure:beta2}).

\begin{figure}[htb]
 \centering
 \begin{subfigure}{0.49\textwidth}
    \includegraphics[scale=0.45]{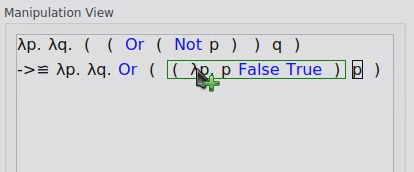}
    \caption{Dropping the argument on the function}
    \label{figure:drop}
 \end{subfigure}
 \begin{subfigure}{0.49\textwidth}
    \includegraphics[scale=0.45]{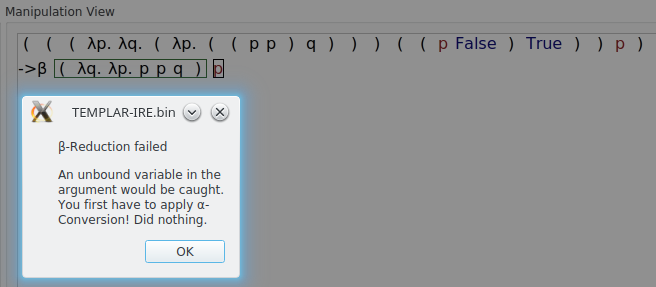}
    \caption{Failing $\beta$-reduction}
    \label{figure:beta-fail}
 \end{subfigure}
 \caption{$\beta$-reduction application}
\end{figure}

Currently, applying the $\beta$-reduction can be performed either by the keyboard
shortcut CTRL+A or dragging the function of the application and dropping it on
the function (see figure \ref{figure:drop}).

\begin{figure}[htb]
 \centering
 \includegraphics[scale=0.6]{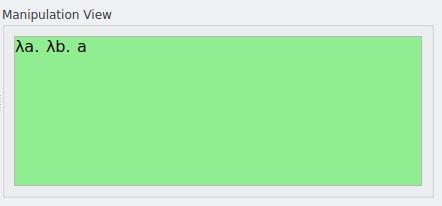}
 \caption{{\lam}-term in normal form}
 \label{figure:normalform}
\end{figure}

The $\beta$-reduction will fail with a warning if it would lead to free variables
becoming bound. Figure \ref{figure:beta-fail} for example shows that the beta
reduction fails as the free variable $p$ (red) would become bound. This way, the
students' awareness for binding scopes shall be increased.  If the term is in
normal form, i.e.~irreducible, the background of the manipulation view is
coloured green (see \ref{figure:normalform}).

\paragraph{$\equiv$-expansion}
The $\equiv$-expansion can be applied on all named term references that have a
valid definition.  It is triggered by double-clicking on the named term
reference. Also, the $\equiv$-expansion is applied automatically to the function
of a reducible application if the $\beta$-reduction is triggered and the function is a
named term reference.

The manipulation view also supports the undo and redo operations, although there are still improvements needed in this functionality as the undo/redo operations need to be synchronised with the editor view.

\section{Challenges}
Concepting a theorem prover as educational tool leads to some problems which do
not have to be considered for pure theorem provers. One of the main problems
evolving from didactic considerations is the necessity of $\alpha$-conversion. In
ordinary theorem proving, no $\alpha$-conversion is needed if a De Bruijn index like
structure is used, as the variable names are not relevant at all.

But when teaching theory of programming, naming clashes are an important aspect
students need to understand. Thus, the $\alpha$-conversion should be supported from the
didactic point of view. Moreover, even if the framework itself does not induce
or have to consider naming clashes, the graphic user interface still may do
this. In the context of our tool, the lambda terms are stored internally with a
De Bruijn like representation. Thus, the $\beta$-conversion and $\equiv$-expansion can
be completely agnostic about variable names. But for readibility reasons, the
variables become relevant as they have to be shown in the text editor panel and
are also parsed from the raw text. A solution would be to rename variables on
clashes on the fly but then again the didactical gain for students would be
smaller as they can just expect the framework to correct their mistakes. Thus,
the user interface needs to address $\alpha$-conversion and the framework itself must
handle naming collisions.

Another challenge in the conception of the tool was the usability, especially
the speed of reaction. As waiting for the result of a $\beta$-conversion is not what
students expect, the rendering speed of the newly generated lambda term needed
to be improved. The main problem of complexity was that on applying a rule to
the current lambda term, the complete term with all reduction, conversion and
expansion steps was rendered. With big terms, this could take several
seconds. Thus, the manipulation view was adopted to only render the newly
created term. But as both the text editor and the manipulation view support
undo/redo, and must be held consistent, they had to share some
information. Thus, the lambda terms were embedded in a document structure that
is used both by the editor and manipulation view.

\section{Supported Platforms and Future Work}

Our tool is implemented in C++ and is available as binary distribution for
Linux. An AppImage for Linux is also available. This bundle contains all
dependencies to work with the tool also on older Linux distributions. A binary
distribution for Microsoft Windows is currently under development
and could already be tested successfully in preliminary versions.  All versions
of the tool are portable, i.e.~no installation is needed.  A distribution as
online tool is under consideration as the
emscripten\footnote{\url{https://github.com/kripken/emscripten}} plugin for the
C++ compiler supports the translation from C++ to JavaScript.
The tool in all its
variants is available at \url{http://www.cs.uni-potsdam.de/~mafrank/} .

There are many possible small improvements. First of all, a stronger decoupling
of the graphic user interface from the framework is planned. The framework
should just apply the rules without checking for potentially caught
variables. The graphic user interface itself should check for those
collisions. Also, the application of the alpha conversion should be improved by
replacing the interactive renaming window with an inline edit mode. Furthermore,
the visualisation of the drag and drop mechanism for $\beta$-reduction will be
improved.  There are still some inconsistencies in the undo/redo functionality
of the manipulation view.

Also, extensions of the tool are planned. This includes a graph view for lambda
terms and improvements on the manipulation view. Lifting the tool to the typed
lambda calculus is also work in progress. Moreover, the current representation
of the source code as file will be replaced by a database with plain text
export.

\section{Conclusion}

We presented an educational tool for undergraduate students that has the aim to
improve their understanding of functional programming while giving them at least
some of the features of state of the art developement environments. Though this
tool is work in progress it was successfully used by multiple students in first
year undergraduate studies. A first release where application was not yet
possible with drag and drop received positive feedback but also some proposals
for improvement.  Much of the functionality was motivated by this feedback as is
also the future work.  We hope that this kind of educational tools will enable
the students to get used to formal tools like proof assistants more easily.

\end{document}